\documentclass{article}

\usepackage[margin=3.5cm]{geometry}
\usepackage{amsmath,amssymb}
\usepackage{graphicx}
\usepackage{authblk}
\usepackage{cite}
\usepackage{hyperref}
\usepackage{titlesec}
\usepackage{setspace}
\usepackage{caption}
\usepackage{booktabs}

\usepackage{siunitx}
\usepackage[capitalise]{cleveref}

\usepackage{command_definitions}

\title{Interference Fringe Mitigation in Short-Delay Self-Heterodyne Laser Phase Noise Measurements}

\author[1]{Jasper Riebesehl\thanks{janri@dtu.dk}}
\author[2]{David C. Nak}
\author[1]{Darko Zibar}

\affil[1]{\small Department of Electrical and Photonics Engineering, Technical University of Denmark (DTU), DK-2800 Kgs. Lyngby, Denmark}
\affil[2]{\small Institute for Quantum Physics, Universität Hamburg, 22761 Hamburg, Germany}

\date{(Dated: May 21, 2025)}

\begin{document}

\maketitle

\begin{abstract}
Self-heterodyne techniques are widely used for laser phase noise characterization due to their simple experimental setup and the removed need for a reference laser. However, when investigating low-noise lasers, optical delay paths shorter than the laser coherence length become necessary. This introduces interference patterns that distort the measured phase noise spectrum.
To compensate for this distortion, we introduce a robust data-driven digital signal processing routine that integrates a kernel-based regression model into a phase noise power spectral density (PN-PSD) equalization framework.
Unlike conventional compensation methods that rely on simplified phase noise models, our approach automatically adapts to arbitrary laser lineshapes by using Kernel Ridge Regression with automatic hyperparameter optimization.
This approach effectively removes the interference artifacts and provides accurate PN-PSD estimates. 
We demonstrate the method's accuracy and effectiveness through simulations and via experimental measurements of two distinct low-noise lasers.
The method's applicability to a broad range of lasers, minimal hardware requirements, and improved accuracy make this approach ideal for improving routine phase noise characterizations.
\end{abstract}

\section{Introduction}
Ultra-stable lasers are critical components in many scientific and engineering applications. Fields such as precision metrology \cite{schioppoComparingUltrastableLasers2022}, quantum optics \cite{dayLimitsAtomicQubit2022}, interferometric sensing \cite{cahillaneLaserFrequencyNoise2021, lihachevLownoiseFrequencyagilePhotonic2022, huangReviewLowNoiseBroadband2023}, and optical communication \cite{colavolpeImpactPhaseNoise2011, jainPracticalContinuousvariableQuantum2022} rely on narrow-linewidth lasers, making the characterization of laser noise performance essential for predicting system behavior and advancing technological boundaries. 

However, measuring phase noise in very low-noise lasers remains a challenging task.
A widely used approach to assess laser phase noise is to generate a beat note on a photodetector by superimposing a highly stable reference laser with the laser under test (LUT)\cite{vonbandelTimedependentLaserLinewidth2016a, zhangCharacterizationElectricalNoise2016}. While this method is straightforward, it necessitates a reference laser whose phase noise is negligible compared to that of the LUT and that operates at a similar wavelength — a combination that is difficult to achieve in practice. In some cases, stabilized frequency combs are used as broadband references or for transferring phase stability across wavelengths\cite{scharnhorstHighbandwidthTransferPhase2015}; however, their complexity, high cost, and limited availability often restrict their utility in routine phase noise characterization.

More sophisticated approaches involve using cross-correlation techniques to generate beat notes with reference lasers \cite{yuanCorrelatedSelfheterodyneMethod2022}. Although these methods relax the stringent requirement on the reference laser’s phase noise, they introduce additional experimental complexity.

Another common strategy employs frequency discriminators \cite{hrabinaFrequencyNoiseProperties2013, yamoahRobustKHzlinewidthDistributed2019, Lin:12}. Ultra-stable reference cavities convert phase fluctuations into amplitude fluctuations; with knowledge of the cavity’s transfer function, the amplitude fluctuations are converted back to recover the phase noise signal. This technique, however, inherently mixes amplitude and phase noise and limits the analysis bandwidth to the cavity linewidth. Additionally, constructing an ultra-stable cavity to enable accurate low-frequency analysis further escalates the experimental complexity.

In contrast, the self-heterodyne method uses the LUT as its own reference, thereby eliminating the need for an external low-noise reference. Typically, the laser output is split into two arms of an interferometer: one arm is frequency-shifted, while the other is temporally delayed. The purpose of the delay is to induce phase decorrelation between the two arms. When the delay exceeds the coherence length of the LUT, the arms become nearly uncorrelated, producing a beat note that mimics that from two independent but identical lasers \cite{moslehiNoisePowerSpectra1986}.

In many practical scenarios, however, a delay arm longer than the laser coherence length is unfeasible. For instance, a laser with a linewidth of \SI{100}{Hz} has a coherence length on the order of \SI{950}{\kilo\meter}, requiring delays of several thousands of \si{\kilo\meter} for complete decorrelation. In practice, long delays can only be realized with fiber coils, and losses become significant when operating outside the transparency window of standard single-mode fibers.
Moreover, thermal fluctuations in long fiber delays translate into excess laser phase noise \cite{hilwegLimitsProspectsLongbaseline2022}.
To address these limitations, short-delay self-heterodyne (SDSH) methods have been developed \cite{armstrongTheoryInterferometricAnalysis}. These methods intentionally use a delay shorter than the coherence length, which sidesteps the need for impractically long delay lines. The drawback of the SDSH approach is that the partial correlation between the arms gives rise to interference fringes at the detector that distort the measured phase noise spectrum, thereby necessitating sophisticated digital post-processing\cite{vanexterExcessPhaseNoise1992}.

Conventional DSP for SDSH involves a spectrum equalization step designed to correct for the interference effects. Often, these methods do not adequately account for detection noise, leading to significant artifacts in the corrected phase noise spectrum \cite{camatelNarrowLinewidthCW2008a, thorndahlthomsenFrequencyNoiseMeasurements2023, ousaidLowPhaseNoise2024, xuLaserPhaseFrequency2015}.
Other techniques correct for interference effects but often assume simple models for the laser line shape \cite{ludvigsenLaserLinewidthMeasurements1998, zhaoNarrowLaserlinewidthMeasurement, huangLaserLinewidthMeasurement2016a, llopisPhaseNoiseMeasurement, fomiryakovNewApproachLaser2021}.

Kantner and Mertenskötter \cite{kantnerAccurateEvaluationSelfheterodyne2023a} proposed an improved approach using a class of power spectrum equalization (PSE) filters to mitigate these artifacts.
While their method is an improvement over the conventional approaches, it relies on simplified model assumptions about the phase noise that are not valid for all laser types and measurement scenarios.

In this work, we extend the PSE filter approach to accommodate lasers with arbitrary lineshapes. We propose a data-driven model that employs Kernel Ridge Regression (KRR) to generate a smooth representation of the measured PSD. This model is non-parametric as it does not require a prior analytical model of the PN-PSD. It is instead learned from an initial suboptimal estimation and then used as a signal-to-noise ratio (SNR) estimator in the PSE framework presented in \cite{kantnerAccurateEvaluationSelfheterodyne2023a}. Crucially, all KRR hyperparameters are automatically selected via cross-validation, removing the need for manual tuning.
This broadens the applicability of the PSE method and improves the accuracy of phase noise estimation across a wider range of laser types.
We demonstrate this enhancement in simulation and on experimental measurements of two seperate low phase noise lasers.
In particular, the lineshapes of the lasers are not easily describable with simple analytical models. Our method removes the interferences fringes close to perfectly from the spectra and produces a more accurate estimate as compared to the conventional method.

Importantly, our experimental setup remains minimal, requiring only a few optical components. By shifting most of the complexity to the digital domain, we achieve a flexible and accessible implementation with relaxed hardware constraints. Moreover, to facilitate reproducibility, we provide a robust implementation of the method.

The remainder of this work is structured as follows: In \cref{sec:setup}, we describe the experimental setup for SDSH along with the digital pre-processing steps required for the measured signal. In \cref{sec:dsp_method}, we detail our data-driven spectrum equalization method and demonstrate its performance using simulated signals. In \cref{sec:exp}, we validate the robustness and effectiveness of our approach with experimental measurements from two different laser systems. Finally, in \cref{sec:conclusion}, we summarize our findings and discuss the limitations of the proposed method.



\section{Short-delay self-heterodyne setup}
\label{sec:setup}
\begin{figure}[t]
    \centering
    \includegraphics[width=0.99\linewidth]{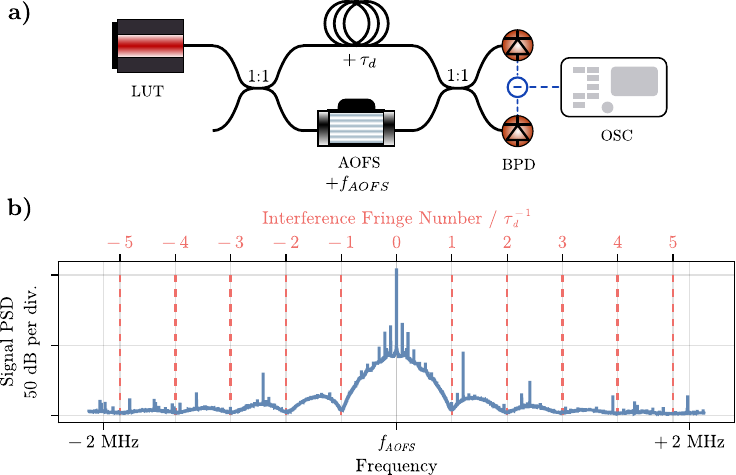}
    \caption{\textbf{a)} Typical experimental setup for short-delay self-heterodyne laser phase noise measurements.
    \textbf{b)} A spectrum of an experimentally detected signal.
    }
    \label{fig:setup}
\end{figure}

The main component of a typical experimental setup for measuring laser phase noise using self-heterodyne detection is an asymmetric Mach-Zehnder interferometer. This setup is illustrated in \cref{fig:setup}. The light from the laser under test (LUT) is divided equally into its two arms. In one arm, the laser frequency is shifted, typicially by few \si{\mega\hertz} to hundreds of \si{\mega\hertz}. This is achieved using an acousto-optic frequency shifter (AOFS), driven by an appropriate signal generator with frequency $f_{\text{AOFS}}$. The other arm features a physical time delay, typically realized as a long coil of fiber.
After the recombination of both arms, the light is detected and digitized using a real-time sampling oscilloscope (OSC). The optical signal is detected with a balanced photodetector (BPD).

Because of the applied frequency shift in one of the arms, the detected signal forms a sinusoidal beat note sinusoid at the difference frequency $f_{\text{AOFS}}$ in the electric domain. 
The digitized time domain signal $y(t_k)$ has the form
\begin{align}
    y(t_k) &= 4\eta \sqrt{P_{LUT}(t_k) P_{LUT}(t_k-\tau_d)} \sin{\left(2\pi f_{\text{AOFS}}t_k + \Delta\phi(t_k)\right)} + \epsilon_y(t_k) \label{eq:raw_signal} \\
    \Delta\phi(t_k) &= \phi(t_k) - \phi(t_k-\tau_d) \label{eq:delta_phi}
\end{align}
where $t_k = k\Delta t$ is the sampled time with sampling interval $\Delta t$,
$\eta$ is a detection proportionality constant, 
$P_{LUT}(t_k)$ is the optical laser power, $\phi(t_k)$ is the instantaneous phase noise of the laser, and $\tau_d$ is the temporal delay introduced in the long interferometer arm.
$\epsilon_y(t_k)$ represents measurement noise originating from laser shot noise and thermo-electric noise in the detector, which is well described as a white Gaussian noise source.
%

In \cref{fig:setup}b), an example of the spectrum of an experimentally detected signal is shown.
The strong central beat note at $f_{\text{AOFS}}$ is flanked by side lobes with equidistant interference fringes.
They are caused by the remaining correlation between $\phi(t_k)$ and $\phi(t_k-\tau_d)$, the difference $\Delta \phi$ of which directly occurs in the detected signal in Eq. \eqref{eq:raw_signal}. The correlation causes constructive and destructive interference, which is manifested in the fringes visible in the spectrum of $y(t_k)$.
The fringes further away from the beat note become less visible as the phase noise drops below the detection measurement noise floor.

\subsection{Phase demodulation}
The phase of the sinusoidal signal $y(t_k)$ can be extracted using the discrete Hilbert transform $\mathcal{H}$.
The complex-valued analytic signal $Y(t_k) = y(t_k) + \mathrm{i} \mathcal{H}[y(t_k)]$ is estimated, from which the phase of the sinusoid can readily be extracted \cite{FELDMAN2001642}. 
Using the expression
\begin{align}
    \Delta\phi(t_k) = \mathrm{unwrap}[\mathrm{arg}[Y(t_k)] - 2\pi f_{\text{AOFS}}t_k] + \epsilon_{\Delta\phi}(t_k),\label{eq:convolution_problem}
\end{align}
a noisy estimate of $\Delta\phi(t_k)$ is obtained. Here, $\mathrm{arg}[\cdot]$ represents taking the complex argument of $Y(t_k)$, which yields the phase estimate. The linear phase evolution is removed by subtracting $2\pi f_{\text{AOFS}}t_k$.
$\mathrm{unwrap}[\cdot]$ describes a phase unwrapping operation that ensures continuity by adding integer multiples of $2\pi$ when sudden jumps (due to phase wrapping) occur \cite{schiemangkAccurateFrequencyNoise2014}.
The noise term $\epsilon_{\Delta\phi}(t_k)$ represents the transformed detection measurement noise $\epsilon_y(t_k)$. It can be assumed to be normally distributed if the beat note is in the center of the signal spectrum \cite{kantnerAccurateEvaluationSelfheterodyne2023a, 2015MeScT..26h5207P}.
%
\section{Data-driven PSD equalization}
\label{sec:dsp_method}
Using Eq. \eqref{eq:convolution_problem}, the differential phase $\Delta\phi$ can readily be extracted from the measurement $y(t_k)$. However, since the path delay $\tau_d$ is shorter than the laser coherence length $\tau_c$ by design, the time auto-correlation function $C_{\phi, \phi}(t_k,t_k-\tau_d)$ does not vanish \cite{ludvigsenLaserLinewidthMeasurements1998}.
Therefore, the functional relation between $\Delta\phi(t_k)$ and $\phi(t_k)$ in Eq. \eqref{eq:delta_phi} can not easily be resolved to extract the actual phase noise $\phi(t_k)$. 

Eq. \eqref{eq:delta_phi} can equivalently be formulated as a convolution:
\begin{align}
    \Delta \phi(t_k) &= \underbrace{(\delta(t_k) - \delta(t_k - \tau_d))}_{\equiv h(t_k)} \ast \, \phi(t_k) + \epsilon(t_k) \label{eq:conv_problem}
\end{align}
where $\delta$ is the Dirac delta distribution, and we omit the subscript of $\epsilon_{\Delta\phi}$.
$h$ is the time-domain of the transfer function of the Mach-Zehnder interferometer in the experimental setup.
In this form, extracting $\phi(t_k)$ reduces to solving a deconvolution problem.

To characterize the phase noise of a laser, we ultimately are interested in its PN-PSD $\SPhi$, from which key quantities such as linewidth can be calculated \cite{didomenicoSimpleApproachRelation2010}.
Formulating Eq. \eqref{eq:conv_problem} in PSD domain by applying the modulus squared of the Fourier transform on both sides yields
\begin{align}
    \SDelPhi = |H(f)|^2 \SPhi + \SEps
    \label{eq:psd_conv_problem}
\end{align}
where $H(f) = 1-e^{i 2\pi f\tau_d}$ is the Fourier transform of $h$.
In Fourier domain, the convolution in Eq. \eqref{eq:conv_problem} transforms into a multiplication.
The PSD of the measurement noise $\SEps$ is assumed to be frequency-independent since $\epsilon$ can be considered a normally distributed random variable. The frequency independency is equivalent to a flat measurement noise floor in the spectrum.

\subsection{Conventional Approach}
A widely used approach to estimate $\SPhi$ is a multiplication of Eq. \eqref{eq:psd_conv_problem} by the inverse transfer function
\cite{camatelNarrowLinewidthCW2008a, thorndahlthomsenFrequencyNoiseMeasurements2023, ousaidLowPhaseNoise2024, xuLaserPhaseFrequency2015}:
\begin{align}
    \SPhiInv \equiv \SDelPhi|H(f)|^{-2} = \SPhi + \underbrace{\frac{\SEps}{2(1-\cos(2 \pi f \tau_d))} }_{\equiv \SPerturb}
    \label{eq:inv_filter}
\end{align}
This method accurately estimates $\SPhi$ when $\SPerturb$ is negligible.
However, $\SPerturb$ tends to infinity at offset frequencies $f$ that are integer multiples of $\tau_d^{-1}$, as the denominator of $\SPerturb$ vanishes.
Subsequently, the estimator $\SPhiInv$ exhibits poles at these frequencies.
\begin{figure}[t]
    \centering 
    \includegraphics[width=1.0\textwidth]{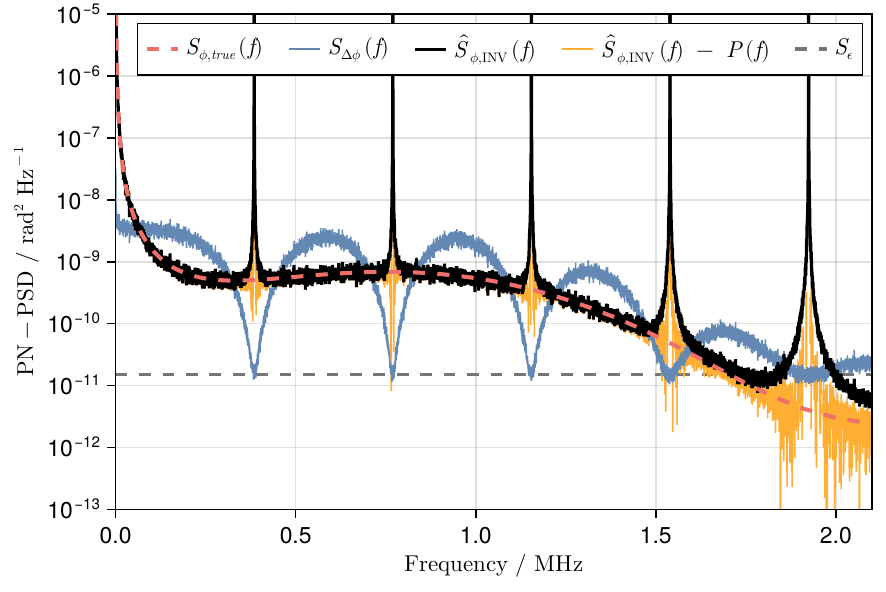}
    \caption{Visualization of the conventional approach to compensate for interference in SDSH measurements. At the interference fringes of $\SDelPhi$, the conventional estimate $\SPhiInv$ diverges away from the true PN-PSD $S_{\phi, true}(f)$.
    }
    \label{fig:krr_method_main_conventional}
\end{figure}

To illustrate this relationship, we simulate a signal in time-domain based on Eq. \eqref{eq:raw_signal}. The simulated shape of the PN-PSD is phenomenologically chosen to resemble a DFB laser with active stabilization. From this arbitrary shape, we use the algorithm in \cite{owensAlgorithmGeneratingFluctuations1978} to generate a time-series realization.
Shown in \cref{fig:krr_method_main_conventional} are the different quantities that occur in Eq. \eqref{eq:inv_filter}.
The shape of $\SDelPhi$ corresponds to a single sideband PSD of $y(t)$, as shown in \cref{fig:setup}b).
The interference fringes at multiples of $\tau_d^{-1}$ are clearly visible, and their minima align with the flat measurement noise floor $\SEps$.

Equalizing the spectrum using the inverse transfer function approach (INV) results in the estimate $\SPhiInv$.
Comparing it to the true phase noise PSD reveals that it produces accurate estimates for frequencies away from the interference fringes.
The estimate deviates from the ground truth at the fringes and when the ratio of $\SPhi$ and $\SEps$ is small.
This can be seen at frequencies $f\gtrapprox\SI{1.7}{\mega\hertz}$, where $\SPhiInv$ diverges from the true phase noise PSD due to the influence of the measurement noise.
The INV estimate can be improved by subtracting $\SPerturb$, according to Eq. \eqref{eq:inv_filter}. This results in better estimates closer to the fringes and the noise floor, as seen in \cref{fig:krr_method_main_conventional}. However, since the estimates are noisy, it produces increased noise and negative values at the poles, which are still present.
In addition, it heavily relies on the underlying assumption of a flat noise floor, and deviations from this assumption will introduce a heavy bias on the estimate. However, this assumption holds in many detection systems.

An improved equalization approach has been presented in \cite{kantnerAccurateEvaluationSelfheterodyne2023a}. Their work considers the presence of measurement noise in Eq. \eqref{eq:psd_conv_problem}, and an optimal filter is derived. Their \textit{power spectrum equalization} (PSE) filter $G_{PSE}$ has the form
\begin{subequations}
\begin{align}
\begin{split}
    \SPhiPse &= |G_{PSE}(f)|^2 \SDelPhi
\end{split} \\
\begin{split}
    |G_{PSE}(f)|^2 &= |H(f)|^{-2} \left(1 + |H(f)|^{-2} \text{SNR}(f)^{-1} \right)^{-1}
\end{split} \\
\begin{split}
    \text{SNR}(f) &= \SPhi \SEps^{-1}. \label{eq:snr_definition}
\end{split}
    \end{align}\label{eq:pse_filter_definition}
\end{subequations} 
This filter mitigates the existence of poles in the spectrum while producing the same results as the inverse filter everywhere else. However, explicitly finding $|G_{PSE}(f)|^2$ presents a challenge as Eq. \eqref{eq:snr_definition} depends on $\SPhi$, which is the quantity we want to estimate.
In \cite{kantnerAccurateEvaluationSelfheterodyne2023a}, this circular dependence is resolved by proposing a phenomenological model of a free-running diode laser for $\SPhi$ and subsequently fitting its parameters.
However, it is generally difficult to find a simple analytical model for the phase noise PSD of an arbitrary laser.
Many types of lasers feature active stabilization loops, which heavily modify the lineshape. In addition, laser phase noise spectra often feature spurious frequencies, which introduce additional complexity and perturb the fitting mechanism.

\subsection{Data-driven approach}
\begin{figure}[t]
    \centering 
    \includegraphics[width=1.0\textwidth]{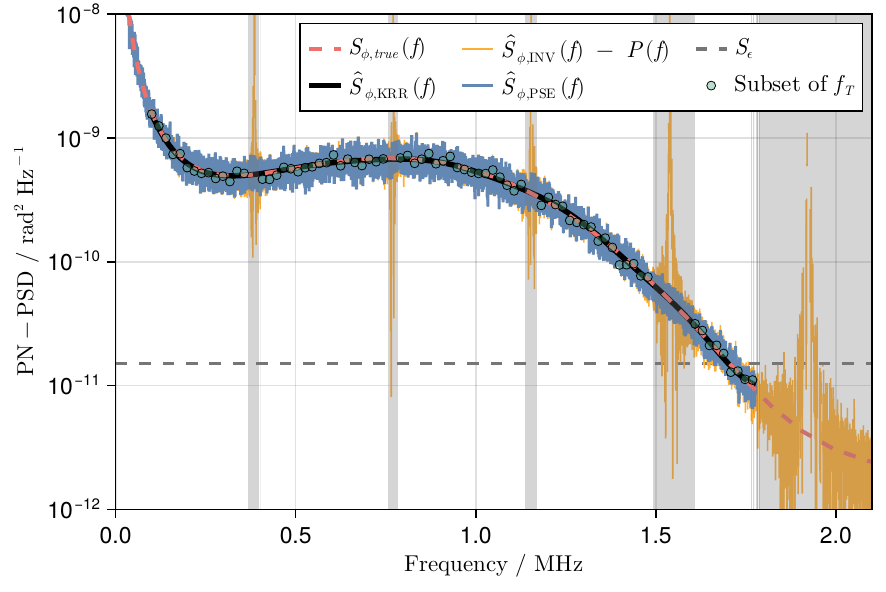}
    \caption{Visualization of the data-driven PN-PSD equalization technique.
    The grey regions indicate the frequency bands excluded from training due to SNR thresholding, here with $T_{\mathrm{SNR}} = 5$ dB. Given a set of training samples $f_T$, the KRR estimate $\SPhiKrr$ almost perfectly overlaps with the true PN-PSD $S_{\phi, true}(f)$, which results in a very accurate final PSE estimate $\SPhiPse$. Notably, $\SPhiPse$ does not diverge at the position of the interference fringes.
    }
    \label{fig:krr_method_main_PSE}
\end{figure}

To resolve these issues and make the PSE filtering approach in Eq. \eqref{eq:pse_filter_definition} feasible for lasers with arbitrary lineshapes, we propose a data-driven approach without explicit assumptions about the laser phase noise.
To approximate $\SPhi$ for the estimation of $\mathrm{SNR}(f)$ in Eq. \eqref{eq:snr_definition}, Kernel Ridge Regression (KRR) is used to learn a model describing $\SPhi$.

\paragraph{Kernel Ridge Regression}
KRR combines ridge regression with the kernel trick to model complex, non-linear relationships in data.
By implicitly mapping the input data into a higher-dimensional feature space using kernel functions, KRR enables efficient modeling beyond simple parametric models. This mapping transforms the non-linear problem into a high-dimensional space, in which it can be solved efficiently. Additionally, a regularization mechanism helps prevent overfitting and improves the model’s generalization ability.
Here we provide a brief overview of KRR. For a more comprehensive introduction, we refer readers to relevant literature on KRR \cite{murphyProbabilisticMachineLearning2022, vuUnderstandingKernelRidge2015}.

Initially, a kernel function $\kappa(f,f')$ is selected that can model the patterns of the data. We use the Radial Basis Function (RBF) kernel for simplicity and its ability to smoothly model non-linear relationships.
It has the form
\begin{align}
    \kappa(f,f') = \exp(-(f - f')^2 / (2\sigma^2))
    \label{eq:kernel_func}
\end{align}
where $\sigma$ is a free parameter, often called the length scale of the kernel.

To train the KRR model, we require a set of training samples $(f^i, S^i)$, where $f^i$ are the frequency values and $S^i$ are the corresponding values of the target function $\SPhi$ to be modeled.
We use the fact that the inverse filtering estimate in Eq. \eqref{eq:inv_filter} is an accurate approximation of $\SPhi$ for frequencies where $\SPerturb$ is negligible compared to $\SPhi$. 
Therefore, we can define a criterion by imposing an SNR threshold as
\begin{align}
    f_\mathrm{T} = \lbrace f \, \mid \, \SDelPhi > T_{\mathrm{SNR}} \SEps \rbrace
\end{align}
where the threshold $T_{\mathrm{SNR}}$ is a free parameter. For a given threshold, this set contains the training frequencies $f_\mathrm{T}$ for which $\SPerturb \ll \SPhi$.
At these frequencies, $\SPhiInvT \approxeq \SPhiT$ according to Eq. \eqref{eq:inv_filter}. Therefore, the training samples to fit $\SPhi$ with KRR can be drawn from $\SPhiInvT$.

The training samples then effectively span the full frequency range, starting from a chosen minimal frequency $f_{KRR,\min} < \tau_d^{-1}$ just below the first pole up to a maximum frequency $f_{KRR,\max}$ where the measurement begins to be limited by the SNR of the measurement.
At the poles, the training set has gaps.
This is visualized in \cref{fig:krr_method_main_PSE}, where the regions with low SNR are shaded in grey.

The objective of KRR then is the smooth interpolation between the gaps, without requiring an explicit model. While the sensitivity to the true phase noise of the experimental setup is fundamentally still zero exactly at the poles, this approach allows for an improved estimation close to the poles and offers a natural continuation of the PSD curve.

The choice of $T_{\mathrm{SNR}}$ influences the maximum measurable frequency offset $f_{KRR, \max}$ of the final phase noise PSD estimate.
A low value will increase the $f_{KRR, \max}$ but increases the chance of the fitting routine to fail due to increased noise. In practice, we have found that values around $T_{\mathrm{SNR}} \approx 2 \approx 3\text{dB}$ strike a good balance.
If the assumption for a flat measurement noise floor holds well, then $T_{\mathrm{SNR}}$ can be reduced, and the training samples can be drawn from $\SPhiInvT - \SPerturbT$. In principle, even values below one are possible.


\paragraph{KRR training}
For the training process, the samples are organized into the input vector $\mathbf{f} = [f^1, f^2, \dots, f^{N_{train}}]^T$ and the target vector $\mathbf{S} = [S^1, S^2, \dots, S^{N_{train}}]^T$ where $N_{train}$ is the number of training samples.
Training the model is then performed by numerically solving the linear equation
\begin{align}
    \mathbf{S} = (\mathbf{K} + \lambda \mathbf{I})\, \boldsymbol{\alpha} \quad \Longleftrightarrow  \quad
    \boldsymbol{\alpha} = (\mathbf{K} + \lambda \mathbf{I})^{-1}\, \mathbf{S} \label{eq:krr_eq}
\end{align}
where $\mathbf{K}_{ij} = \kappa(f^i, f^j)$ is the kernel matrix, $\lambda$ is the regularization parameter, $\mathbf{I}$ is the identity matrix and $\boldsymbol{\alpha}$ is a vector of dual coefficients.
$\boldsymbol{\alpha}$ then represents the coefficients of the trained KRR model.

To improve numerical stability, the target training samples $\mathbf{S}$ are logarithmically scaled before solving \eqref{eq:krr_eq}.
Using the trained model, the target function value for any new input \(f'\) can be inferred as
\begin{align}
    \hat{S}_{\phi,KRR}(f') = \sum_i \alpha_i \kappa(f', f^i).
    \label{eq:KRR_estimation}
\end{align}
Using Eq. \eqref{eq:KRR_estimation} allows for the evaluation of the KRR model at any frequency $f'$, in particular at the poles where no training samples are available. This allows the interpolation from a limited number of training samples to a continuous curve.
The KRR estimate for the simulated signal is shown in \cref{fig:krr_method_main_PSE}.

The method's ability to model non-linear behavior is limited by the choice of kernel function. Only smooth and slowly changing behavior can be modeled when choosing the RBF kernel. However, for our application this assumption is justified.
Additional limitations of KRR are its $\mathcal{O}(N_{train}^3)$ computational complexity scaling with $N_{train}$ and sensitivity to outliers, which can be addressed by carefully selecting the training samples.
To reduce the computational complexity of the KRR fit, a random subset in the order of a few hundred of frequencies is selected from $f_\mathrm{T}$ to create a set of training samples.

\paragraph{Hyperparameter optimization}
\begin{figure}[t]
    \centering
    \includegraphics[width=1.0\linewidth]{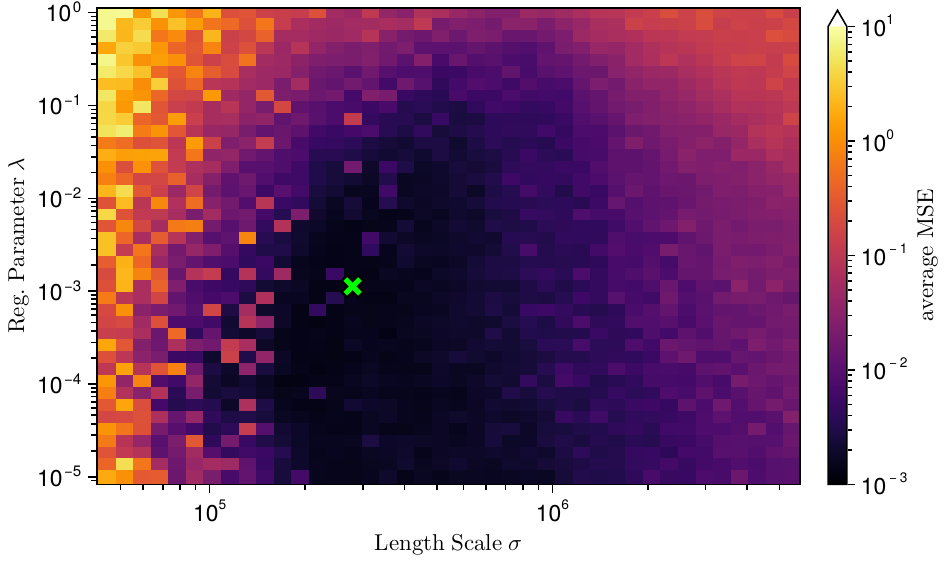}
    \caption{Hyperparameter grid search using grouped n-fold cross-validation, where $N_{train}=600$, $n=8$ and $G=25$. The green cross indicates the optimal set of hyperparameters. As the KRR is performed using logarithmic scaling, the MSE is calculated in logarithmic scaling as well.}
    \label{fig:x_val_grid}
\end{figure}
To find the optimal regularization parameter $\lambda$ and the kernel function length scale $\sigma$ which occurr in Eqs. \eqref{eq:kernel_func} and \eqref{eq:krr_eq} respectively, we perform grouped $n$-fold cross-validation \cite{murphyProbabilisticMachineLearning2022}. First, the set of training samples is segmented into groups of $G$ frequency-consecutive samples. These groups are then randomly assigned to $n$ folds containing the same amount of groups.
In each iteration of the cross-validation, the samples in one of the folds are used as validation data, while the samples in the other $(n-1)$ folds are used as training data. A KRR model is trained using the training set and a given pair of hyperparameters $(\lambda, \sigma)$. The KRR model is then used to predict the target values $\mathbf{S}_{val}$ for the samples in the validation fold. The validation mean squared error (MSE) is calculated as
\begin{align}
    \mathrm{MSE}_{val} = 1/N_{val} \sum_i (S^i_{val} - \hat{S}_{\phi,KRR}(f^{i}_{val}))^2
\end{align}
where $N_{val}$ is the number of validation samples. The MSE serves as a score function to determine the quality of the KRR model fit.
This is repeated until each fold has been the validation fold once. The average MSE over all $n$ folds serves as an indication for the generality of the KRR model, given a set of hyperparameters.

To find the optimal hyperparameters, the cross-validation is repeated for different combinations of hyperparameters. For simplicity and since the model has only two hyperparameters, a grid search is performed. For each pair, we perform the cross-validation and obtain the average MSE.

In \cref{fig:x_val_grid}, the average MSE for different hyperparameter combinations are shown for the simulated measurement shown in \cref{fig:krr_method_main_PSE}.
A hyperparameter set with minimal error is clearly discernible.
The parameter set with the lowest average MSE can then be considered as the optimal parameters for the given dataset. 

\paragraph{Final PN-PSD estimate}
After performing the KRR fit, the estimate $\SPhiKrr$is obtained from Eq. \eqref{eq:KRR_estimation}.
Since the training samples are not drawn from the full frequency range, the KRR estimate is only meaningful between the minimum and maximum KRR frequencies $f_{KRR, \min/\max}$. Therefore, a joint estimate is defined as 
\begin{align}
    \SPhiJoint = \left\{
        \begin{array}{llr}
        \SPhiKrr & \text{if} & f_{KRR, \min} < f < f_{KRR, \max}\\
        \SPhiInv & \text{if} & f < f_{KRR, \min}
    \end{array}
\right.
\end{align}
to cover the full frequency range. For frequencies below $f_{KRR, \min}$, we copy the estimate $\SPhiInv$, which is justified as no interference fringes occur below $f_{KRR, \min}$ by construction.
This joint estimate for $\SPhi$ is finally used in Eq. \eqref{eq:snr_definition} to calculate the SNR, and, subsequently, the PSE filter estimate $\SPhiPse$ using Eq. \eqref{eq:pse_filter_definition}.
This final estimate is also shown in \cref{fig:krr_method_main_PSE}. The poles are completely mitigated, and the estimate follows the ground truth $S_{\phi, true}(f)$.

\subsection{Choice of optical path delay length}
The discussed equalization methods require a precise estimate of the experimental time delay $\tau_d$, as it appears in Eqs. (\ref{eq:inv_filter},~\ref{eq:pse_filter_definition}).
Incorrect values for $\tau_d$ result in skewed estimates around the poles, and low-frequency values of the PSD are scaled incorrectly.

A data-driven estimation of $\tau_d$ can be performed by numerically finding the pole frequency difference in $\SDelPhi$. The mean inverse of the frequency difference corresponds to $\tau_d$, according to Eq. \eqref{eq:psd_conv_problem}. However, this does require that at least one, preferably a few, poles are visible in $\SDelPhi$.
For poles to be visible, they have to occur at frequencies where the phase noise is larger than the measurement noise. This ultimately depends on the amplitude of the beat note, the detection measurement noise floor, and the level of phase noise at the pole \cite{zhaoInfluenceNoiseFloor2022}.
An example of this relation can be seen in \cref{fig:setup}b), where only a few fringes are visible. Since the phase noise of this laser is low at high offset frequencies, it is not visible below the detection noise floor.

Increasing the delay pushes the interference fringes closer to the beat note, which will cause more of them to be visible. This usually improves the estimation accuracy of $\tau_d$ since more averaging can be performed.
On the other hand, the delay should not be unnecessarily long to avoid loss and additional induced phase noise through vibrations in the delay fiber.
In practice, the ideal delay needs to be found experimentally.

\section{Application to experimental measurements}
\label{sec:exp}
To show the method's effectiveness in different experimental settings, we apply it to measurements of two different lasers. Both measurements were taken in the same experimental setup equivalent to the one illustrated in \cref{fig:setup}a).
The delay in this setup was realized using a fiber delay of circa \SI{530}{\meter}, which causes interference fringes at multiples of \SI{375}{\kilo\hertz}.
The exact delay is estimated using a numerical peak fitting routine on the poles in the spectrum of $\SDelPhi$.
The KRR-based method is then applied to each measurement independently to calculate and apply the PSE filter.

\begin{figure}[t]
    \centering
    \includegraphics[width=1.0\linewidth]{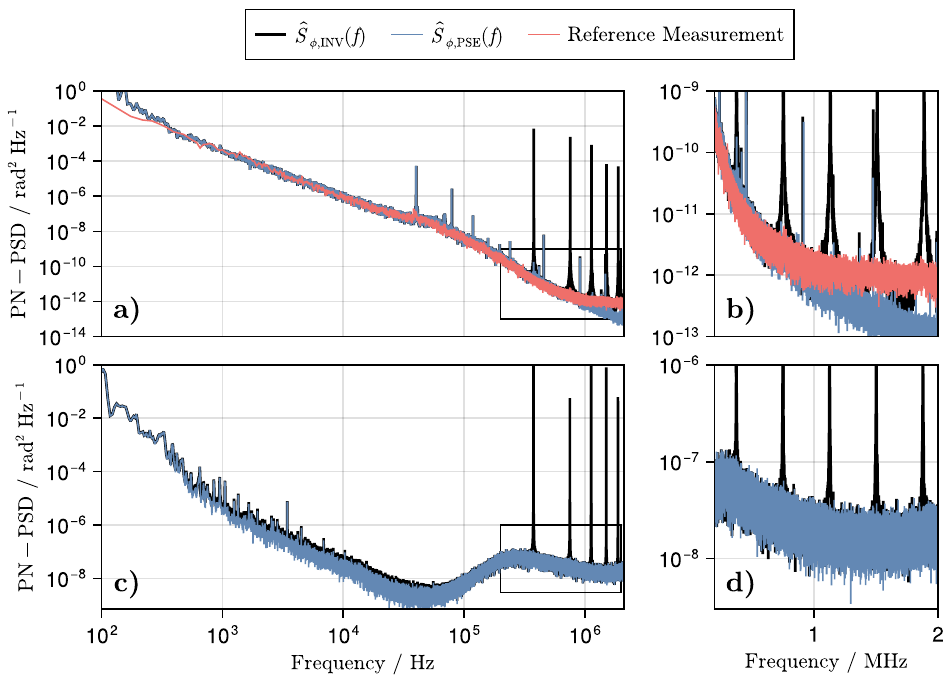}
    \caption{Application of the equalization method to experimental measurements. \textbf{a)} Phase noise PSD estimates of a commercial fiber laser. \textbf{b)} zoom of marked region in a).
    \textbf{c)} Phase noise PSD estimates of a custom-built, externally stabilized DFB laser. 
    \textbf{d)} zoom of marked region in c).\\
    }
    \label{fig:exp}
\end{figure}

The results for both lasers are shown in \cref{fig:exp}.
In a-b), the phase noise PSD of a commercial fiber laser source (\textit{NKT Photonics Koheras BASIK E15}) at $\sim$\SI{1550}{\nano\meter} is shown. Both the simple inverse transfer function estimate $\SPhiInv$ as well as the data-driven PSE estimate $\SPhiPse$ are shown.
In addition, an independent phase noise measurement is shown, which was recorded in a regular heterodyne setup with two identical copies of the fiber laser. This type of measurement does not require any spectrum equalization, which makes it a suitable reference measurement.
The reference measurement agrees closely with both the INV and PSE estimates for frequencies below the first interference fringe. Apart from spurious frequencies likely caused by detector nonlinearity or quantization noise, all curves have very good overlap.
In \cref{fig:exp}b), the estimates at the poles can be seen in detail.
At the first two fringes, the PSE estimates agree well with the reference while $\SPhiInv$ diverges. At higher frequencies, the reference is measurement noise limited. However, the fringes that occur in the INV estimate are effectively removed in the PSE estimate.

The second example is the phase noise measurement of a custom-built, externally stabilized distributed feedback (DFB) laser at $\sim$\SI{2}{\micro\meter} \cite{nak2025fibercoupledexternalcavitydfblaser}.
No reference measurement is available for this laser, as no suitable low-noise local oscillator laser for a heterodyne measurement was accessible.
In \cref{fig:exp}c), we again observe the agreement of both equalization methods, apart from the pole artifacts. The zoomed-in view in d) shows the perfect mitigation of the fringe artifacts and a naturally smooth continuation of the curve.
For this laser in particular, finding an analytical model with few fitable parameters would be very difficult. The lineshape at the fringe positions is dominated by the servo bump of the active feedback loop.
While no ground truth is available in this experimental measurement, the simulation study, as well as the previous experimental example, indicate that $\SPhiPse$ is a good estimate for the true phase noise PSD $\SPhi$.

\section{Conclusion}
\label{sec:conclusion}
Using a simulation and two experimental measurements, we have shown how KRR-based power spectrum equalization can significantly improve phase noise PSD estimates in SDSH setups. In particular, the discussed method can effectively mitigate interference artifacts caused by the coherence of the light in the experimental setup. The method is very general there are no strict requirements on the laser lineshape, which makes it applicable to a wide range of lasers.

The main limitation of the method is the interpolation capability of KRR. In our approach, we used a simple kernel to reduce complexity. However, due to this simplification, the method typically has difficulties modeling sharp features in the PSD. In future work, more advanced modeling approaches (such as Gaussian processes or neural networks) can be used to improve upon this aspect.

\subsection*{Funding}
This work has been funded by the SPOC Center (Grant No.
DNRF 123), the Villum Fonden (Grant No. VI-POPCOM 54486) and by the Deutsche Forschungsgemeinschaft (DFG, German Research
Foundation) (Grant No. HE 2334/15-2).

\subsection*{Disclosures}
The authors declare no conflicts of interest.

\subsection*{Data Availability Statement}
The data and code that support the results of this work are available upon reasonable request.

\bibliographystyle{unsrt}
\bibliography{main}

\end{document}